\documentclass[letterpaper,dvipsnames]{article}
\pdfoutput=1 % if your are submitting a pdflatex (i.e. if you have
\usepackage{jheppub} % for details on the use of the package, please
\usepackage{hyperref}

\usepackage{amsmath}
\usepackage[T1]{fontenc} % if needed
\usepackage{braket}
\usepackage{subcaption}
\usepackage{soul}
\usepackage{array}
\usepackage{xcolor}
\usepackage{cancel}
\usepackage{slashed}
\usepackage{comment}
\usepackage[many]{tcolorbox}    	% for COLORED BOXES (tikz and xcolor included)

\newtcolorbox{boxA}{
    fontupper = \bf,
    boxrule = 1.5pt,
    colframe = black % frame color
}

\newtcolorbox{boxE}{
    enhanced, % for a fancier setting,
    boxrule = 0pt, % clearing the default rule
    borderline = {0.75pt}{0pt}{main}, % outer line
    borderline = {0.75pt}{2pt}{sub} % inner line
}

\usepackage[compat=1.1.0]{tikz-feynman}

\author{Vazha Loladze,}
\author{Arthur Platschorre}
\author{and Mario Reig}

\affiliation{Rudolf Peierls Centre for Theoretical Physics, 
University of Oxford, Parks Road, Oxford OX1 3PU, United Kingdom}
\emailAdd{vazha.loladze@physics.ox.ac.uk}
\emailAdd{arthur.platschorre@physics.ox.ac.uk}
\emailAdd{mario.reiglopez@physics.ox.ac.uk}

\title{Higher Axion Strings}

\abstract{We study the minimal requirements to obtain axion strings for axions with exponentially good quality. These ingredients appear in theories where an axion coming from a higher-form gauge field mixes with the phase of a complex scalar field in a situation that resembles higher-groups. The resulting axion is perturbatively massless and inherits a high-quality shift symmetry from the global higher-form symmetry while being compatible with a post-inflationary axion scenario. Due to differences and resemblances with both, extra-dimensional and field theory axions, we call this field the \textit{higher axion}. To this end, we study a toy model on a 5-dimensional manifold with boundary. The boundary hosts the complex scalar that provides axion strings through standard mechanisms.
In addition, we study how these scenarios may arise in heterotic string theory and type~II string compactifications.}

\begin{document} 

\maketitle

\section{Introduction}
There is an ever-growing experimental program \cite{Graham:2015ouw,Irastorza:2018dyq} dedicated to finding the QCD axion. This particle could provide an elegant resolution to the strong CP problem \cite{Weinberg:1977ma,Wilczek:1977pj,Peccei:1977hh}, could account for all of the observed dark matter \cite{Preskill:1982cy,Dine:1982ah,Abbott:1982af} and is ubiquitous in string theory compactifications \cite{Svrcek:2006yi,Arvanitaki:2009fg}. As this particle would be feebly coupled to us, a large fraction of the experimental efforts aim to detect the QCD axion through smaller-scale resonant searches, making the axion mass one of the most important quantities for current searches. The axion mass parameter space these experiments have to cover is vast and seems impractical unless more theoretical guidance is provided. 

In theory, the mass of the axion can be inferred from the observed dark matter (DM) abundance if the late-time axion density is to saturate this abundance. In practice, this inference has proven  difficult due to the large hierarchy of scales between the axion decay constant and relevant cosmological parameters. Large theoretical and numerical efforts are being dedicated to reducing the uncertainties in this inference in the so-called post-inflation axion DM scenario -- that is, the scenario where the Peccei-Quinn (PQ) shift symmetry of the axion is spontaneously broken after the end of inflation \cite{Borsanyi:2016ksw,Vaquero:2018tib,Gorghetto:2018myk,Buschmann:2019icd,Gorghetto:2020qws,Kim:2024dtq,Saikawa:2024bta,Correia:2024cpk,Benabou:2024msj}. In this scenario, a network of axion strings is expected to form and simulations have shown that they can dominate the production of DM axions. Because of this, the required decay constant is smaller than a situation with only misalignment contribution and the axion mass prediction increases. Recent simulations point to a QCD axion decay constant around the scale $f_a\sim 10^{10}-10^{11}$ GeV~\cite{Saikawa:2024bta,Correia:2024cpk,Benabou:2024msj}.

This post-inflationary scenario where the axion comes from the phase of a complex scalar fields benefits from theoretical simplicity and is very predictive, but is plagued by extreme sensitivities to UV physics dubbed the QCD axion quality problem \cite{Kamionkowski:1992mf}. This problem challenges us to explain the absence of local operators breaking the anomalous PQ symmetry, which are expected to be present both in a generic effective field theory and due to quantum gravitational effects~\cite{Kallosh:1995hi}. Standard arguments for the QCD axion require such (misaligned) breaking sources to have either extremely suppressed couplings or to be generated at a high order in operators (for a recent review see~\cite{Dine:2022mjw}). 

Many theoretical efforts have been dedicated to alleviating the axion quality problem. Broadly speaking, efforts in four-dimensional field theory have mainly focused on recovering the axion shift symmetry as an accidental symmetry of the particle  content of the theory. This scenario corresponds to a  category of axions in which dangerous higher-dimensional operators are forbidden by the gauge charges of the particle content. Examples include models with additional gauge redundancies like $\mathbb{Z}_{N}$ \cite{Chun:1992bn,Ringwald:2015dsf} or $U(1)$ (see for example \cite{Fukuda:2017ylt}), and composite axion models 
from the confinement of additional gauge groups \cite{Randall:1992ut,Lillard:2017cwx,Lillard:2018fdt,DiLuzio:2017tjx,Gavela:2018paw,Contino:2021ayn,Vecchi:2021shj,Choi:2022fha,Babu:2024udi,Babu:2024qzb}
While the latter models are able to explain the large separation between the PQ scale and weak scale as dynamically generated, these models require involved model building and the post-inflationary scenario is, in some cases, disfavored by heavy relic production.

Akin to other fine-tuning problem of the Standard Model, an alternative to four-dimensional symmetries can be played by higher dimensional constructions \cite{Arkani-Hamed:1998jmv,Randall:1999ee}. Extra-dimensional field theories or string theory constructions are generically expected to contain a large number of axions depending on the number of closed cycles and gauge fields in the compact dimensions, the so-called \textit{axiverse} \cite{Arvanitaki:2009fg}. In these scenarios, axions arise as the zero modes of a bulk higher-form gauge fields integrated over closed extra-dimensional cycles and inherit a high-quality shift symmetry from the more robust global higher-form shift symmetries of gauge fields \cite{Craig:2024dnl,Agrawal:2024ejr}, ameliorating several quality issues that plague its four-dimensional counterparts. See \cite{Reece:2024wrn} for a recent, comprehensive review.

However, this category of axions lacks from the predictability of the post-inflationary field theoretic axion. Intuitively, 
the axion in such scenarios comes from the Goldstone of the spontaneously broken higher-form symmetry. In contrast to its four-dimensional counterparts, in the standard cosmological history this symmetry is never expected to be restored and therefore no phase transition or network of solitonic axion strings is expected. In this sense, higher-form axions generically correspond to the pre-inflationary scenario, where the DM abundance depends quadratically on the initial misalignment angle, reducing the predictability of the theory~\cite{Benabou:2023npn} and inference of the axion mass.  

In this work we present a different kind of axions where both, the predictivity of the post-inflationary scenario as well as the exponentially good axion quality, can be present simultaneously. To gain some intuition let us consider a simple field theoretic model where the axion comes as a linear combination of the zero-mode of the higher-dimensional gauge field and the phase of a complex scalar.  An illustrative example is an orbifold scenario with 5d bulk gauge fields and a 4d boundary -- that is, a $3-$brane localised at the orbifold fixed point -- where the complex scalar lives\footnote{Another interesting 5d model with a distinct set-up  was presented in \cite{Cheng:2001ys} also offering a high-quality axion. }. At low energies, the model has a one-dimensional approximate vacuum manifold on the brane: 
\begin{equation}\label{eq:U(1)_mixing}
    U(1)_{PQ} \equiv \frac{U(1)^{(0)} \times U(1)^{(1 \downarrow 0)}}{U(1)_{\mathrm{gauge}}}\,.
\end{equation}   
The $U(1)^{(0)}$ symmetry is an ordinary $0$-form symmetry associated with the phase of a complex scalar on the brane. The $U(1)^{(1 \downarrow 0)}$ symmetry is a $0$-form symmetry that 
that descends from a global 1-form symmetry in the 5d bulk. A diagonal direction on the vacuum manifold is a gauge redundancy, leaving the orthogonal direction $U(1)_{PQ}$ as the axion, $\theta$. Any symmetry breaking or lifting of this vacuum manifold needs to -- by construction -- leave the gauge direction invariant and therefore 
always has to break the more robust $U(1)^{(1 \downarrow 0)}$ symmetry, guaranteeing an exponentially good quality. 
This situation differs qualitatively from field theory as well as higher-dimensional axions, and resembles situations that appear in the context of higher-group symmetries \cite{Cordova:2018cvg}. For this reason, we call the 4d axion, $\theta$, a \textit{higher axion}. 
 
The approximate vacuum manifold also allows for string solutions -- that is, $1$-dimensional field configurations that are topologically protected by the property that when a non-contractible loops in physical space is traversed, the field configurations traverses a non-contractible loop on the vacuum manifold. There are two independent global string solutions, corresponding to the $U(1)^{(1 \downarrow 0)}$ and $U(1)^{(0)}$ non-contractible cycles, that can be related by a local string involving both cycles. The $U(1)^{(1\downarrow 0)}$ global strings admit no symmetry restoration and are expected to require a decompactified  core \cite{March-Russell:2021zfq} and are generically difficult to be produced in the standard cosmological history (see however \cite{Benabou:2023npn,Cline:2024vbd}). On the other hand, the $U(1)^{(0)}$ strings can be produced by the Kibble-Zurek mechanism \cite{Kibble:1976sj} as the temperature of the universe drops below the spontaneous breaking scale of the symmetry. At this instance, the gauged direction on the vacuum manifold provides mass to a gauge boson, which -- in the scenario of a small bulk -- is much larger than the other scales associated with the $U(1)^{(0)}$ strings.
In this limit, the cosmology is similar to that of ordinary field theory strings.

As noted above, the key ingredient is the mixing between the higher-form and a zero-form symmetry. This mixing is also the crucial ingredient in \textit{pseudo-}anomalous $U(1)$ scenarios where 4d gauge anomalies are canceled by bulk higher-form fields. This intuition allows us to identify string theory constructions where similar mechanisms operate and provide \textit{higher axions} -- that is a pertubatively massless Goldstone boson that emerges as a linear combination of a higher-form axion and an ordinary 0-form axion (phase of a complex scalar). We will sketch how to obtain these axions in heterotic string theory as well as in type IIB compactifications.

The structure of the paper is as follows. The set-up and details of the toy model are described in section \ref{section2} along with the quality of the resulting higher axion. Section \ref{section3} describes the string solutions and the cosmological history. In section \ref{sec:string} we briefly describe the minimal ingredients to realise this mechanism in string theory. Finally, in section \ref{discussion} we summarise our findings.  

\section{The minimal set-up: 5d theory with boundary}
\label{section2}

Let us consider a standard orbifold scenario consisting of a $5$-dimensional space-time with four flat dimensions parametrized by coordinates $x^{\mu}$ and one compact dimension of length $R$, parametrized by an angular coordinate $\phi \in [0,2 \pi)$ or $y = R \phi$. The angular coordinate is subject to the gauge identification $(x^{\mu},\phi) \sim (x^{\mu},-\phi)$, effectively reducing the compact space to $S^{1}/\mathbb{Z}_{2}$ \cite{Kawamura:1999nj}. This orbifold has two fixed points, $\phi = 0 , \pi$, which can host $(3+1)$-dimensional field theories on fixed $3$-branes\footnote{We will focus on the physics on one of the branes. In principle, the discussion can be generalised to include fields on both branes, 
but this would not alter the conclusions regarding the low-energy physics.}. Latin indices will denote $5$D indices $M=0,1,2,3,5$, whilst greek indices will be used for the four flat dimensions $\mu = 0,1,2,3$. 

The bulk of the $5$-dimensional spacetime contains a gauge field $A_{M}$ with coupling strength $g_A$. Under the orbifold parity identification, the gauge field transforms as $(A_{\mu},A_{5}) \sim (-A_{\mu},A_{5})$. Additionally, the gauge field has an ordinary gauge identification
\begin{equation}
    A \rightarrow A + d \lambda_{A}\,. 
\end{equation}

In the limit where this gauge field is decoupled from all other dynamics of the theory, there exists a 
global one-form and two-form symmetries, the electric $U(1)^{(1)}$ and magnetic $U(1)^{(2)}$, under which non-contractible Wilson lines and ‘t Hooft surface operators transform by a $U(1)$ phase, respectively \cite{Bhardwaj:2023kri}. These symmetries are spontaneously broken and act non-linearly  on their respective Goldstone bosons, the photon $A_{M}$ and the dual photon $\widetilde{A}_{MN}$, by shifting by a closed one (two)-form,
\begin{align}
A &\xrightarrow{U(1)^{(1)}} A + c^{(1)} , \quad dc^{(1)} = 0  \,,
\\
\tilde{A} & \xrightarrow{U(1)^{(2)}} \tilde{A} + \tilde{c}^{(2)}  , \quad d\tilde{c}^{(2)} = 0  \,.
\end{align}
The electric (magnetic) symmetry are emergent below the scale of the lowest electrically charged particles (magnetically charged lines), at which point electric (magnetic) field lines (surfaces) cannot end and their number penetrating any surface is conserved. It is well-established that this conservation is only broken by non-perturbative Schwinger-like pair production processes, providing the exponential protection for these symmetries.    
As described around Eq.~\eqref{eq:U(1)_mixing}, part of the light axion will be identified with the zero mode of this gauge field along the compact dimension, called $a$, and provided by the Wilson loop around the compact dimension
\begin{equation}
    a = \oint A \ dy\,.
\end{equation}
On the compact dimension, the $1$-form shift symmetry of the bulk gauge field reduces to a $0$-form shift symmetry $U(1)^{(1\downarrow 0)}$ of the zero mode
\begin{equation}
    a \rightarrow a + c, \qquad dc = 0 \,.
\end{equation}
This symmetry inherits the exponentially good quality from the $1$-form shift symmetry of the gauge field, $A$. In this scenario, the axion is an Aharanov-Bohm phase around an extra-dimensional circle. Only particle worldlines that wrap this circle are sensitive to this phase, $a$. These worldlines are  suppressed by the exponential of the action for such events, which scales with the size of the extra dimension and provides the exponential protection. This is akin to virtual Schwinger pair production processes that break this symmetry are now associated with particle-anti-particle pairs that are created and loop around the extra-circular dimension where they annihilate.

In addition to the bulk gauge field, we assume that the branes at the orbifold fixed point $y =0$ supports another gauge field $C_{\mu}$ with gauge identification 
\begin{equation}
    C \rightarrow C + d\lambda_{C}\,.
\end{equation}

A crucial ingredient of our mechanism is that the shift symmetry of the zero mode $a$ will transform under this identification. One way to accomplish this compatible with locality\footnote{Generically, a $\delta$-function localisation in shifts of $1$-form gauge fields can be inherited from a  $1$-form analogue of the Green-Schwarz mechanism \cite{Green:1984sg}. For example, the gauge field $A$ can shift as
$A\rightarrow A + \lambda_Cd\alpha $ under $U(1)_C$ gauge transformations where $\alpha \sim \alpha + 2\pi$ is a $0$-form periodic field. In the presence of a background flux, $d\alpha = \delta(x)$, the gauge transformation becomes localised, although we stress that our model is independent of a particular UV realisation. This is analogous to the Green-Schwarz mechanism where a 2-form field $B$ shifts as $B \rightarrow B + \lambda_{C} d\tilde{A}$ where $\tilde{A}$ is a $1$-form gauge field. This scenario is generic in string theory compactifications as discussed in section \ref{sec:string}, where background fluxes $d\tilde{A} = \delta (S_{2})$ on a sphere $S^{2}$ localise a gauge symmetry on a cycle or D-brane. The modern terminology for a GS mechanism is that of a (gauged) higher group \cite{Cordova:2018cvg}, further motivating the name higher axions. } is 
\begin{equation}
    A_{5} \rightarrow A_{5} + \delta(y)  \lambda_{C} + \partial_{5} \lambda_{A}\,.
    \label{gauged1}
\end{equation}

As the gauge interaction \eqref{gauged1} is local in position space, the entire KK tower shifts under the gauge symmetry:
\begin{equation}
    a \rightarrow a + \lambda_{C}\,, \qquad  A_{5}^{(n)} \rightarrow A_{5}^{(n)} +  \lambda_{C} - n\lambda^{(n)}_{A}\,, \qquad n = 1,2,\dots 
\end{equation}

The zero mode is part of a tower of Kaluza-Klein (KK) modes $(A_{\mu}^{(n)},A_{5}^{(n)})$ provided by the Fourier transform of $A_{M}$ along the compact dimension subject to the orbifold identification
\begin{equation}
    A_{\mu}(x,\phi) = \frac{1}{\pi}\sum_{n=1}^{\infty}A_{\mu}^{(n)}(x) \sin{\left(n\phi\right)}\,, \qquad  A_{5}(x,\phi) = \frac{a(x)}{2\pi R} + \frac{1}{\pi R}\sum_{n=1}^{\infty}A_{5}^{(n)}(x) \cos{\left(n\phi\right)\,.} 
\end{equation}

The five-dimensional action on the orbifold has to be invariant under the gauge identification. When reduced into the individual KK components, the four-dimensional Lagrangian is: %\MR{check units.. is $g$ dimensionful?} 
\begin{equation}
    \mathcal{L} \supset -\frac{R}{4g_A^{2} \pi} \sum_{n =1 }^{\infty} \left( \partial_{\mu}A^{(n)}_{\nu} - \partial_{\nu} A^{(n)}_{\mu}\right)^{2} + \frac{1}{2g_A^{2} \pi R} \sum_{n = 1}^{\infty}\left( \partial_{\mu}A_{5}^{(n)} - n  A_{\mu}^{(n)}  -C_{\mu} \right)^{2} + \frac{1}{4g_5^{2} \pi R} \left(\partial_{\mu} a - C_{\mu} \right)^{2}\,.
    \label{Lagrangian}
\end{equation}
The physical consequences of having mixed the gauge redundancies between the bulk gauge field and brane gauge field are most apparent in the the axial gauge choice $A_{5}^{(n>0)} = 0$. In this gauge, all gauge bosons obtain a mass apart from one zero eigenvector\footnote{The zero eigenvector of the mass matrix is  
$C_{\mu} - \sum_{n=1}^{\infty} \frac{e^{2}R}{n g_5^{2} \pi} A_{\mu}^{(n)}$ but will henceforth be simply called $C_{\mu}$} of the mass matrix. The additional tower of massive gauge fields is heavy $m^{2}_{KK} \sim  \frac{1}{R^{2}}$ and can be integrated out. Upon integrating out these gauge bosons, the electric coupling of $C_{\mu}$ shifts by 
\begin{equation}
    \frac{1}{e^{2}} \rightarrow \frac{1}{e^{2}} \left(1 + \frac{e^{2}R\pi}{6g_5^{2}} \right)\,.
\end{equation}

As seen in Eq.~\eqref{Lagrangian}, the would-be higher dimensional axion $a$ has been eaten by the 4d gauge boson $C_\mu$. To introduce an axion in the 4d EFT we have to modify the boundary theory at the orbifold fixed point $y = 0$. Let us consider a brane-localised charged complex scalar $\Phi=|\Phi|e^{ib}$, whose phase $b$ shifts under the gauge transformations of $C_{\mu}$ as
\begin{equation}
    b \rightarrow b + \lambda_{C}\,.
\end{equation}
The potential of this scalar forces it to have a minimum at the vev $|\Phi| = F_{\Phi}$. Below this scale, the low energy Lagrangian is of the form:
\begin{equation}
        \mathcal{L} \supset \frac{1}{4e^{2}} \sum_{n =1 }^{\infty} \left( \partial_{\mu}C_{\nu} - \partial_{\nu} C_{\mu}\right)^{2} +\frac{F_{a}^{2}}{2} \left(\partial_{\mu} a - C_{\mu} \right)^{2} + \frac{F_{\Phi}^{2}}{2} \left(\partial_{\mu} b - C_{\mu} \right)^{2} - V(a-b) \,,
        \label{twolagrangian}
\end{equation}
where $F_{a}^{2} = \frac{1}{2\pi Rg_A^{2}}$ is the decay constant of the zero mode $a$, which is fixed and necessarily tied to the compactification scale, and we have included a potential for the un-gauged linear combination $a-b$, whose  contributions will be detailed in section \ref{axionquality}.

An alternative construction would be to consider $U(1)_C$ as a bulk gauge symmetry. In this case, one imposes the orbifold boundary condition $(C_\mu,C_5)\rightarrow (C_\mu,-C_5)$. The IR limit of this theory will coincide with the one described above. 
In both cases, we will be interested in coupling the un-gauged axion linear combination, $a-b$, to the SM gauge bosons. Due to the gauge transformation of $a$ and $b$, this will induce boundary-localised anomalies that have to be canceled by a bulk Chern-Simons (CS) interaction. 
Anomaly cancellation will work in a qualitatively different way in these two cases as we will study later in more detail.

\subsection{The Axion}
The Lagrangian \eqref{twolagrangian} below the compactification scale $1/R$ and spontaneous breaking scale $F_{\Phi}$ includes a zero mode $a$ and a phase $b$. The linear combination $a+b$ is `eaten' by the gauge field $C_{\mu}$ and does not appear in the low-energy EFT. The orthogonal combination will form a compact scalar $\theta \sim \theta +2 \pi$ -- the \textit{higher axion} -- with decay constant $F_{\theta}$. This can be seen by diagonalizing \cite{Fraser:2019ojt} to the eaten/un-eaten basis $(F_{\pi} \pi,F_{\theta} \theta)$ by rotating the original basis vector $\left(F_{a}a, F_{\Phi} b \right)$ as
\begin{align}
    F_{\pi} \pi &= \cos{\left(\alpha\right)}F_{a} a + \sin{\left(\alpha \right)}F_{\Phi} b\,, \\ F_{\theta}  \theta &= -\sin{\left(\alpha \right)}F_{a} a + \cos{\left(\alpha \right)}F_{\Phi} b\,. 
    \label{cosines}
\end{align}

Requiring that $\theta$ is uncharged under the gauge transformations fixes the angle as
\begin{equation}
    \cos{\left(\alpha \right)} = \frac{F_{a}}{\sqrt{ F_{a}^{2} +  F_{\Phi}^{2}}} \,,\qquad \sin{\left(\alpha \right)} = \frac{ F_{\Phi}}{\sqrt{F_{a}^{2} +  F_{\Phi}^{2}}}\,. 
\end{equation}
Defining $\pi$ as having charge $1$ under $U(1)_C$ and the periodicity of $\theta$ as $\theta \sim \theta +2 \pi$ lifts the degeneracy between the new decay constants $F_{\pi}$ and $F_{\theta}$ and the fields $\pi$ and $\theta$ and uniquely defines the $\pi$ field as
\begin{equation}
    \pi = \frac{1}{F^{2}_{\pi}} \left(F_{a}^{2}a + F_{\Phi}^{2} b \right)\,, \qquad F_{\pi}^{2} =   F_{a}^{2} +  F_{\Phi}^{2}\,,
\end{equation}
and the axion $\theta$ as
\begin{equation}
    \theta = -a + b \,,\qquad F_{\theta} = \frac{F_{a} F_{\Phi}}{F_{\pi}} \,.
\end{equation}
After performing this rotation, the Lagrangian can be expressed in this basis as
\begin{equation}
    \mathcal{L} \supset \frac{1}{2} F^{2}_{\pi}\left(\partial_{\mu} \pi -  C_{\mu} \right)^{2} +  \frac{1}{2}F^{2}_{\theta} \left( \partial_{\mu} \theta \right)^{2}  - V(\theta)\,.
    \label{axionpion}
\end{equation}
The field $\pi$ will be eaten by the gauge boson $C_{\mu}$, which can be integrated out, leaving behind a perturbatively massless axion $\theta$.

\subsection{Axion Quality}
\label{axionquality}
The axion $\theta$ of the previous section is best described as excitations on the vacuum manifold:
\begin{equation}
    U(1)_{PQ} \equiv \frac{U(1)^{(0)} \times U(1)^{(1 \downarrow 0)}}{U(1)_{\mathrm{gauge}}}\,.
\label{vacuummanifold}
\end{equation}   
The ability of this axion to solve the strong CP problem relies on the quality of its anomalous continuous shift symmetry and its coupling to gluons. For generic axions, this quality can be plagued by several extreme sensitivities to UV physics. In the effective field theory (EFT), any operator consistent with the particle content and symmetries of the theory should generically be present and could generate a potential for the axion $V_{\mathrm{break}}(a)$. Standard arguments require such (misaligned) sources of shift symmetry breaking to be extremely suppressed: 
\begin{equation}\label{eq:Vbreak}
    V_{\mathrm{{\cancel{PQ}}}} < 10^{-10} f_{\pi}^{2} m_{\pi}^{2}\,,
\end{equation}
if the CP violating phase in the strong sector is not finely-tuned. For a generic 4d field theory axion, $\varphi$, this problem is further exacerbated by the expectation that quantum gravity is believed to break all global symmetries \cite{Kallosh:1995hi}. In this case the breaking potential is expected to have a lower bound of \cite{Kamionkowski:1992mf}
\begin{equation}
    V_{\mathrm{{\cancel{PQ}}}} \sim \frac{F^{n}_{\varphi}}{M_{Pl}^{n-4}} e^{i n\varphi} + \mathrm{h.c.}
\end{equation}
where $n$ is the number of axion insertions involved in the breaking. Astrophysical constraints favour axion decay constants $F_{\varphi} \gtrsim 10^{8}$ GeV~\cite{Caputo:2024oqc,Hook:2018dlk}, which in turn requires such gravitational sources to be suppressed so that the dimension of the leading operator is $n \geq 14$ in order to be consistent with \ref{eq:Vbreak}.

It is well-known that extra-dimensional axion scenarios largely circumvent these extreme sensitivities to UV physics \cite{Choi:2003wr}. Any effect breaking the shift symmetries of such axions will require an \textit{extension} into the compact dimension and will be suppressed by the exponential of the action which is proportional to the size of the compact space. In the mechanism described here, any effect lifting the vacuum manifold \eqref{vacuummanifold} of the axion $\theta$  by construction requires to leave the gauge direction invariant and  always has to break the more robust $U(1)^{(1 \downarrow 0)}$ symmetry, guaranteeing an exponentially good quality analogous to other extra-dimensional axion scenarios. 

In this case, the least suppressed gauge invariant operator that breaks the shift-symmetry of the un-eaten linear combination, $\theta = -a + b$ is
\begin{equation}
    \mathcal{O} = c \Lambda_{UV}^3\Phi^* e^{-S}e^{ia} + \text{h.c.}\,,
\end{equation}
with $c\sim O(1)$. This operator leads to the axion potential
\begin{equation}
   V(\theta)= -c\, F_\Phi \Lambda_{UV}^3 e^{-S}\cos \theta\,.
\end{equation}
We notice, as expected for higher-dimensional axions, that $\theta$ has exponentially good quality despite that it contains, in part, the phase of the complex scalar field, $\Phi$. In typical higher-dimensional and string constructions the instanton action is related to the 4d effective gauge coupling of bulk symmetries as $S\sim \frac{2\pi R}{g_5^2}\sim \frac{2\pi }{\alpha (R^{-1})}$. Provided that the gauge couplings are small in the UV, this guarantees the quality of $\theta$ to solve the strong CP problem.

\subsection{Anomaly cancellation and axion couplings in 4d}

In general, the axion has a topological coupling to gluons as required by the PQ solution to the strong CP problem and to electromagnetism, which provides the main detection strategies in a large class of experiments \cite{Irastorza:2018dyq}. In the extra-dimensional axion scenario, this coupling is facilitated by extending the gauge fields of electromagnetism and QCD as zero modes of extra-dimensional gauge fields with field strengths $F^{EM},F^{QCD}$~\cite{Choi:2003wr}. This extension into the five-dimensional bulk allows for 
CS interactions between the Standard Model gauge fields and bulk gauge field,
\begin{equation}\label{eq:bulk_axion_coupling}
    \mathcal{L} \supset   \frac{\kappa_{EM}}{16 \pi^{2}} A_{M}  \left(F_{NL} \widetilde{F}^{MNL} \right)_{EM} + \frac{\kappa_{QCD}}{16 \pi^{2}} A_{M}  \left(F_{NL} \widetilde{F}^{MNL} \right)_{QCD}\,,
\end{equation}
and the coupling constants $\kappa_{EM},\kappa_{QCD} \in \mathbb{Z}$ are to be integer valued in the normalization that the electric coupling constants sit in front of the gauge field kinetic terms. 

Due to the gauge transformation Eq.~\eqref{gauged1}, these CS terms require additional fermions localised on the brane that cancel the induced anomalies with respect to the gauge redundancies of $C_{\mu}$. The brane fermions admit couplings to the complex scalar $\Phi$, at which point the fermions become massive below the spontaneous breaking scale and can be integrated out, resulting in a low-energy four-dimensional coupling of the axion $\theta$ to electromagnetism and QCD,
\begin{equation}\label{eq:brane_loc_axion_coupling}
    \mathcal{L} \supset   \frac{\kappa_{EM}}{16 \pi^{2}} \theta  \left(F \widetilde{F} \right)_{EM} + \frac{\kappa_{QCD}}{16 \pi^{2}} \theta  \left(F \widetilde{F} \right)_{QCD}\,,
\end{equation}
similar to standard four-dimensional QCD axions. Altogether, Eqs.~\ref{eq:bulk_axion_coupling} and \ref{eq:brane_loc_axion_coupling}   result in the cancellation of the mixed anomalies $[SU(3)_{\rm QCD}]^2\times U(1)_C$ and $[U(1)_{EM}]^2\times U(1)_C$.

The fermions on the brane introduce additional  cubic anomalies $[U(1)_C]^3$. In the minimal version of the mechanism where $U(1)_C$ is a brane-localised gauge symmetry the cubic anomaly cannot be directly canceled by a CS term since $C_\mu$ does not propagate into the bulk. These anomalies are canceled by a subtle mechanism which is a lower-dimensional analogue of M-theory \cite{Horava:1996ma}. In order to comply with gauge invariance of $C_{\mu}$, the Bianchi identity of the field strength $F^A$ of the bulk gauge field $A$ is altered on the brane 
\begin{equation}
    dF^A = \delta(y) F^{C} \wedge dy\,.
\end{equation}
Similar to M-theory, the Bianchi identity can be solved by having the field strength along the direction of the brane given by 
\begin{equation}
    F^A_{\mu \nu} = \frac{1}{2}\epsilon(y) F^{C}_{\mu \nu} + \dots
    \label{expansionF}
\end{equation}
where $\epsilon(y)$ is the parity-odd step function around the orbifold fixed point $y = 0$. The dots indicate terms that have to vanish on the brane due to the orbifold parity identification.
This altered field strength admits a  CS self-coupling in the bulk
\begin{equation}
   \mathcal{L} \supset \frac{\kappa_{A}}{8 \pi^{2}} A \wedge  F^A\wedge {F}^{A} \,.
\label{selfchernsimons}
\end{equation}
Plugging the expansion \eqref{expansionF} back into this CS coupling cancels the boundary-localised cubic anomaly $[U(1)_C]^3$ for appropriately chosen $\kappa_{A} \in \mathbb{Z}$, which corresponds to the anomaly coefficient induced by chiral fermions on the boundary. 
 
\subsubsection{Alternative model with anomaly cancellation from bulk CS term}
If the $U(1)_C$ gauge field is allowed to propagate in the bulk we can impose the boundary condition $(C_\mu,C_5)\rightarrow (C_\mu,-C_5)$ under orbifold parity. In this case, the 4d cubic anomaly can be directly canceled by a CS term of the type
\begin{equation}
  \frac{\kappa_C}{8\pi^2}  A\wedge F^C\wedge F^C\,.
\end{equation}
The cancelation of the anomaly is guaranteed, as before, by the $A_M$ bulk gauge field  transformation
\begin{equation}
    A_5\rightarrow A_5+\delta(y)  \lambda_{C} + \partial_{5} \lambda_{A}\,.
\end{equation}

While the previous mechanism that we discussed in section~\ref{section2} is more similar in spirit to M-theory, this mechanism described here is closer to how anomalies are cancelled in some type II string theory models. The field $A_M$ plays the role of RR fields in type II string theory and QCD, EM, and $U(1)_C$ will be the gauge sectors realised on the worldvolume of $D-$branes. The anomalies localised on the world-volume of the $D-$branes, in this case $[U(1)_C]^3$ in 4-dimensions, are canceled by charge inflow from the gravitational bulk, since $A_M$ behaves as an equivalent RR field.

\section{Higher Axion Strings}
\label{section3}
In this section we argue that there exist string solutions that can be thermally produced through a standard Kibble-Zurek mechanism \cite{Kibble:1976sj}. In the limit that the mass of the gauge boson, $C_\mu$, is much larger than the VEV of the complex scalar, $|\Phi|$, the string consist mostly of the un-eaten axion mode and its decay will contribute to the relic abundance of axion dark matter. 
 
The vacuum manifold \eqref{vacuummanifold} is said to admit string solutions; field configurations with $2$-dimensional worldsheets that are topologically protected by the non-trivial winding in the $U(1) \times U(1)$ vacuum manifold as one traverses a loop in spacetime enclosing the string. The topological index associated with the winding is best classified in terms of the winding numbers of the original phases $a$ and $b$ around the string. 

The phenomenology of strings with non-trivial windings of $a$ is that of Stückelberg strings. There is no symmetry restoration at the core of the string, but instead local effective field theory is expected to break down. In principle, special inflationary paradigms exist to UV complete such a description and non-thermally create these strings including \cite{Benabou:2023npn,Cline:2024vbd}  which always require a phase in the early universe where the 4d EFT breaks.  However, if formed, such strings likely overproduce the DM abundance in scenarios without strong warping. 

Strings in which the phase $b$ winds do admit a description of the core captured by the four-dimensional field theory. In the cosmological history of the universe, such strings are expected to form when the temperature drops below the scale $|\Phi|=F_{\Phi}$. Below this scale, the angle $b$ obtains random values in uncorrelated patches and a network of strings forms due to the Kibble-Zurek mechanism. Regions in space-time with non-trivial windings of $b$ will enclose string cores. In the absence of any potential for $b$, the string solution that winds once is axially symmetric and the complex scalar and massive gauge field at a distance $r$ and angle $\alpha$ in a plane orthogonal to the string is described by: 
\begin{equation}
    \Phi(r,\alpha) = f(r) e^{i \alpha} \,,\qquad C_{\alpha}(r,\alpha) = \frac{g(r)}{r}\,,
\end{equation}
where $f(r)$ and $g(r)$ have to vanish sufficiently quickly near the core of the string. 

The full string solution including the core profile can be found by minimizing the string's tension or energy per unit length \cite{Niu:2023khv}. The tension of the global string is logarithmically dominated by the IR dynamics and  proportional to the scale of symmetry breaking $F_{\Phi}^{2}$. However, a small winding of the massive gauge field around the string further reduces this scale to $F_{\theta}$. The massive gauge field 
obtains the IR value: 
\begin{equation}
    C_{\alpha} = \frac{1}{r}\frac{F_{\Phi}^{2}}{F_{a}^{2}+F_{\Phi}^{2}} \,,
\end{equation}
implying that these strings carry a small amount (non-integer) of flux. The full tension, including that of the core, can then be estimated as
\begin{equation}
    T \sim \frac{\pi F_{\Phi}^{4}}{2 F_{\pi}^{2}} + \pi F_{\Phi}^{2} \ln{\frac{m}{eF_{\pi}}} + \pi  F^2_{\theta} \ln{\frac{ e
    F_{\pi}L}{2}} + \frac{\pi}{2} F_{\Phi}^{2}\,.
\end{equation}
Where $m$ is the mass of the radial mode of $\Phi$ and size of the string core. The length scale $L$ provides an IR cut-off of the theory, which could be the diameter of a single closed string or the distance between two extended strings of opposite orientation.  

In the limit $F_{a} \gg F_{\Phi}$, it is disfavored to have any heavy gauge field $C_{\mu}$ around the string and we recover the ordinary tension of a global string
\begin{equation}
    F_{a} \rightarrow F_{\pi} \qquad F_{\Phi} \rightarrow F_{\theta} \qquad T \rightarrow \frac{\pi}{2} F_{\theta}^{2} + \pi F_{\theta}^{2} \ln{\frac{mL}{2}}\,.
\end{equation}
In this limit, the string profile is that of a global string and we recover an ordinary axion string around which the \textit{higher axion} $\theta(x)$ that winds by $2 \pi$. Their evolution and  contribution to the relic axion abundance will follow standard axion string simulations \cite{Borsanyi:2016ksw,Vaquero:2018tib,Gorghetto:2018myk,Buschmann:2019icd,Gorghetto:2020qws,Saikawa:2024bta,Kim:2024dtq,Correia:2024cpk,Benabou:2024msj}.

\section{Heterotic \& Type II Axion Strings }\label{sec:string}
The field theoretic models above resemble anomaly cancellation in different string theories, inspiring us to build string theory realisations, although we stress that none of the features of the toy model in section~\ref{section2} rely on any specific string theory embedding. 

The $D=10$ dimensional low energy bosonic excitations common to all superstring theories are captured by the dilaton $\phi$, the anti-symmetric two-index Kalb-Ramond field $B_{MN}$ and the graviton $g_{MN}$, and collected in the 
$\mathcal{N}=1$ SUGRA action \cite{Polchinski:1998rr},
\begin{equation}
    S = \frac{1}{2 \kappa_{0}^{2}} \int d^{10}x \ \sqrt{-g} e^{-2\phi} \left(R - \frac{1}{2} H_{M N L} H^{MNL} +4 \partial_{M}\phi \partial^{M}\phi \right)\,.
\end{equation}
In addition, the bosonic part of the various types of superstring theories contain additional forms and gauge fields specific to the type of string theory under consideration, see \cite{Ibanez:2012zz} for a comprehensive discussion on the different possibilities. 

Typically, the vacuum state of the ten-dimensional geometry is $M_{4} \times K$, where $K$ is a Calabi-Yau if the theory admits $\mathcal{N}=1$ SUSY in $D=4$~\cite{Candelas:1985en}. The low-energy four dimensional spectrum is determined by the topological invariants of the Calabi-Yau.

The $D=4$ dimensional massless spectrum includes the graviton and a set of massless chiral superfields $S,T^i,Z^j$. The superfield $S$ is the dilaton-axion chiral superfield, whose real part is the dilaton zero mode $s$ and complex part is the model-independent axion that is dual to the zero-mode of $B_{\mu\nu}$. The superfield $T^i$ is the set of Kähler moduli, whose real part $t_{i}$ sets the size of the $2$-cycles (whose number is determined by the Hodge number $h^{1,1}$) and complex part are the model-dependent axions corresponding to the integral of $B_{MN}$ along the cycle. The volume $V$ of the Calabi-Yau can be expressed in terms of the Kähler moduli $t_{i}$ and the triple intersection numbers $d_{ijk}$ as,
\begin{equation}\label{eq:CY_volume}
    \kappa = \frac{V}{6} = d_{ijk} t^{i} t^{j} t^{k}\,.
\end{equation}
The additional four-dimensional graviton and the complex structure moduli $Z^j$ are not relevant for the rest of our discussion. 

In the case of heterotic string theory, there will in addition be complex scalars field $\Phi_{i}$ descending as zero modes of the $D=10$ gauge fields in the presence of non-trivial fluxes. The low-energy gauge group will also contain Abelian gauge groups $U(1)_{j}$ with gauge fields $V_{j}$ with the complex scalars carrying gauge charges $p_{ij}$. 
In type II, fluxes and $D-$branes on the Calabi-Yau may play a relevant role and result in an additional set of chiral superfields $\Phi_{i}$ with charges $p_{ij}$ under a set of (distinct) $U(1)_{j}$ gauge groups with gauge fields $V_{j}$. 

If the $U(1)_{j}$ symmetries are anomalous from the $D=4$ dimensional perspective, then (part of) the moduli should also transform under the symmetry to cancel the anomaly by a standard Green-Schwarz mechanism~\cite{Green:1984sg}. In general, this transformation of the moduli can be a complicated  transformation under each of the $U(1)$ symmetries. In line with the toy model, the low energy axion will be a combination of the phase of the chiral superfields $\Phi_{i}$ and the imaginary parts of the moduli fields. 

The $D=4$, $\mathcal{N}=1$ supersymmetric Lagrangian contains terms of the form,
\begin{equation}
    \mathcal{L} \supset \int d^{4}\theta \ \left[\Phi_{i}^{\dagger} e^{-p_{ij}V_{j}} \Phi_{i} + K(S,T^i,V_j) \right] + \int d^{2}\theta \frac{1}{4} f(S,T^i) W^{j}W_{j} \,.
\end{equation}
The Kähler potential $K$ is a function of only $S+\overline{S}$ and $T^i+\overline{T^i}$ up to non-perturbative corrections~\cite{Polchinski:1998rr}. The Kähler potential also includes the gauge fields $V_{j}$, in order to compensate for the transformation of $S$ and $T^i$ under the gauge symmetry. These symmetries can be anomalous, and the anomaly is canceled by the shift of the gauge kinetic function $f(S,T^i)$ under the symmetry, which is coupled to the super field strength $W^{j}$ of the low energy gauge fields. For the model-independent axion, the Kähler potential in the absence of gauge fields is of the universal form $K(S+\overline{S}) = -M_{P}^{2} \ln{\left(S+\overline{S}\right)}$. 

The effect of the Kähler potential is two-fold. It contributes to the mass of the gauge field $V_{j}$ as,
\begin{equation}
    M^{2}_{j} = \frac{\partial^{2} K}{\partial V^{2}_{j} } \,,
    \label{eq:stringymassgauge}
\end{equation}
and to the $D$-term as a Fayet-Iliopoulos (FI) term,
\begin{equation}\label{eq:general_d-term}
    D_{j} =  \frac{\partial K}{\partial V_{j}} -p_{ij} |\Phi_{i}|^{2} \,.
\end{equation}
In order to preserve $N=1$ supersymmetry, these $D$ terms have to vanish \cite{Dine:1987xk}. In a general scenario, Eq.~\eqref{eq:general_d-term} this fixes the VEV of the complex scalar $\Phi_{i}$ to be close to the mass of the gauge boson \cite{Dine:1987xk}.  However, in principle, the $D$ terms and mass $M_j^{2}$ can be scaled independently and there exists special loci in moduli space along $\frac{\partial K}{\partial V_{j}}$. In this scenario, the VEV of the complex scalars can be kept suitably small by small deviation from this locus whilst the mass of the gauge boson $M_j^{2}$ is typically proportional to the string scale, $M_s$. 

\subsection{Heterotic $E_{8} \times E_{8}$}
In heterotic string theory~\cite{Gross:1984dd}, the massless $10$ dimensional spectrum consist of $\mathcal{N}=1$ supergravity with either an $SO(32)$ or $E_{8} \times E_{8}$ supersymmetric Yang-Mills gauge group and contains an universal dilaton and model-independent axion $S =s + i \sqrt{2} \sigma$ and a number of Kähler moduli $T^{i} = t^{i} + 2i \chi^{i}$. The Kähler potential in the absence of gauge fields is,
\begin{equation}
    K = - M_P^2 \ln{\left(S + \overline{S} \right)} -  M_P^2\ln{\kappa}  \,.
\end{equation}
where $\kappa$ is defined in Eq.~\ref{eq:CY_volume}. While the field metric on the model-independent axion is universal, the field metric on the Kähler moduli is,
\begin{equation}
    G_{ij} = - \frac{3}{2} M_P^2 \left( \frac{\kappa_{ij}}{\kappa} - \frac{3}{2} \frac{\kappa_{i} \kappa_{j}}{\kappa^{2}}\right) \,,\qquad \text{with: }\,\,\, \kappa_{i} = d_{ijk} t^{j} t^{k} \,, \qquad \kappa_{ij} = d_{ijk} t^{k} \,.
\end{equation}
We will be focussing on Calabi-Yau manifolds with fluxes in an $S(U(1)^{5})$ subgroup embedded in $E_{8}$ as $S(U(1)^{5}) \supset SU(5) \supset E_{8}$, corresponding to the line-bundle models considered in \cite{Blumenhagen:2005ga,Anderson:2011ns,Anderson:2012yf}. This set-up can result in low energy axion with decay constants -- that is, below the string-scale for certain values of $\kappa_i$ and $U(1)$ charges $q_{ij}$, as originally shown in  \cite{Buchbinder:2014qca}. Other axion set-ups with low decay constants in heterotic string theory and M-theory can be found in \cite{Im:2019cnl}. 

In certain constructions, the low-energy theory is the  GUT gauge group $SU(5)_{\rm GUT} \times S(U(1)^{5})$. The fact that the $5$ symmetries are really $4$ is encoded by identifying the charge vectors $\vec{q}_{i} = (q_{ij})$ and $\vec{\widetilde{q_{i}}} = (\widetilde{q}_{ij})$ if,
\begin{equation}
    \vec{q}_{i}-\vec{\widetilde{q_{i}}} = \mathbb{Z}(1,1,1,1,1)  \,.
\end{equation} 
%For realistic compactifications, the volume of the $2$-cycles is large $t^{i} \gg 1$, but small compared to the dilaton $t^{i}/s \ll 1$.
In addition, the charged matter spectrum contains several complex scalar fields $\Phi_{i}=\hat{\Phi}_ie^{i b_i}$ with charges $p_{ij}$ under the $S(U(1)^{5})$ symmetries. When these complex scalars obtain a VEV, the $S(U(1)^{5})$ symmetries are broken and the  fluxes occupy a larger non-Abelian subgroup of rank $4$ inside $SU(5) \supset E_{8}$. The general potential for the matter fields requires a full description of this non-Abelian bundle and the Calabi-Yau, which is far beyond the scope of this work, and would require a combination of the methods presented in \cite{Blesneag:2015pvz,Douglas:2006hz,Harvey:2021oue,Constantin:2024yxh}.  

Instead, our focus will lie on the $D$-terms and the Kähler potential. In general, several of the $S(U(1)^{5}) \cong U(1)^{4}$ symmetries are anomalous, and therefore the Kähler moduli must participate in the anomaly cancellation and transform as,
\begin{equation}
    \chi^{i} \rightarrow \chi^{i} +  q_{ij} \lambda_{j}\,, \qquad  \sigma \rightarrow \sigma + q_{\sigma j} \lambda_{j} \,.
    \label{eq:kählertrans}
\end{equation}
The gauge-invariant Kähler potential is therefore of the form,
\begin{equation}
    K(S + \overline{S} -q_{\sigma j} V_{j}, \  T_{i} + \overline{T}_{i} -q_{i j} V_{j} ) \,.
\end{equation}
The mass of the $j$-th gauge boson can be recovered from the Kähler potential of the Kähler moduli by equation \eqref{eq:stringymassgauge} and transformation \eqref{eq:kählertrans}, 
\begin{equation}
    M^{2}_{j} = \frac{\partial^{2} K}{\partial V_{j}^{2}} = q_{ij} G_{ik} q_{kj} + \frac{q_{\sigma j}^{2}}{4s^{2}} M_P^2 \,.
    \label{eq:massheterotic}
\end{equation}
In this case, the mass is simply proportional to the field space metric on the moduli suitably contracted with the charge vector. Given that the typical 2-cycle volume is smaller than the overall CY volume in string units, $t^{i}/s \ll 1$, it can be appreciated that the dilaton term constitutes a small correction.

The Kähler potential also gives rise to a contribution to the $D$-term, which in combination with contribution of the complex scalar, becomes
\begin{equation}\label{eq:D-term_heterotic}
    D_j =  M_P^2\left ( \frac{3 q_{ij} \kappa_{i}}{\kappa} + \frac{q_{\sigma j}}{s} \right ) -  p_{ij} |\Phi_{i}|^{2}\,.
\end{equation}
The Calabi-Yau admits special loci in moduli space, called split loci, along which the first two terms vanish (see \cite{Anderson:2011ns,Anderson:2012yf} for details). Close to the split loci, the VEV of the complex scalar field $\Phi_{i}$ can be made suitably small whilst preserving supersymmetry and a vanishing $D$-term. 

At these points in moduli space, the low-energy theory is that of a gauge field $V_{j}$ with mass $M_j^2$, a higher-dimensional axion $a_j=\sum_{i} c_{ij} \chi^{i}$, where the coefficients $c_{ij}$ depend on the $U(1)_j$ charge of the $\chi^i$ field, and the phase of the complex scalar, $b_j$. The decay constant of $a_j$ is proportional to the gauge boson mass, $F_a^2\sim M_j^2$, while the VEV of the complex scalar can be suitably smaller, $|\Phi|^2\ll M_j^2=F_a^2$. We remark that the gauge invariant linear combination -- the \textit{higher axion} $\theta$ -- contains  $b_j$ as well as $a_j$ in close analogy the toy model in section~\ref{section2}.

The low-energy theory admits axion strings solutions with a profile for the radial mode and a corresponding winding of its  phase. The  core of the higher axion string does not correspond to a decompactification limit, but rather a change in the nature of the vector bundle on the Calabi-Yau. The geometric interpretation of the radial profile in the $D=10$ dimensional theory is that, far away from the core of the string, the complex scalar has a VEV, and the fluxes occupy a rank $4$ subgroup of $SU(5) \supset E_{8}$. At the core of the string, the complex scalar vanishes, the fluxes sit in an $S(U(1)^{5})$ and the Kähler moduli lie along the split locus. This results in an apparent enhanced symmetry group $S(U(1)^{5})$ at the core of the string, but the gauge fields \textit{eat} the higher-form axion which shifts under the anomalous $U(1)$s as required by anomaly cancellation. 

\subsection{Type II String Theory}
In this section we discuss the minimal ingredients to build models which capture the key aspects of the mechanisms described above in the context of type IIB string theory\footnote{Similar constructions may be envisaged for type IIA.}.
The Kahler potential part neglecting the dilaton is given by
\begin{equation}
    \mathcal{K}_{\rm IIB} = -2\ln \mathcal{V}\,,
\end{equation}
where, as before, the total volume can be written in terms of the triple intersection points, $d_{ijk}$ and the Kahler parameters, $t_i$,
\begin{equation}
    \mathcal{V}=\frac{1}{6}d_{ijk}t^it^jt^k\,.
\end{equation}
In the case of theories where gauge sectors arise from $D7-$branes wrapping 4-cycles it is more convenient to work with the complexified 4-cycle moduli,
\begin{equation}
    T^i=\frac{1}{2} d_{ijk}t^jt^k+i\theta^i\,.
\end{equation}
We define the 4-cycle volumes as
\begin{equation}
    \tau^i=\frac{\partial\mathcal{V}}{\partial t_i}=\frac{1}{2} d_{ijk}t^jt^k\,.
\end{equation}
Inverting these relations to write $t^i$ and $\mathcal{V}$ in terms of 4-cycle volumes in closed form is not always possible. However, we know that the CY volume $\mathcal{V}$ is a homogeneous function of degree $\frac{3}{2}$ in the volumes $\tau^i$.

Let us now describe a toy model that captures the ingredients discussed in section~\ref{section2}. Consider two $D7-$branes wrapped on intersecting 4-cycles denoted as $W_1$ and $W_2$. If these branes were fully coincident, they would provide a $U(2)$ gauge theory in the 4d EFT. However, because of the intersection over a lower-dimensional space, the symmetry is $U(1)_1\times U(1)_2$ \cite{Uranga:933469}. At the intersection there will be chiral matter which is charged under both $U(1)$s.  Among these fields there will be complex scalar fields which we call $\Phi=\hat{\Phi}e^{ib}$ with charges $(+1,-1)$, but more general possibilities could be envisaged with more involved model building. Due to the charges of the chiral fields living at the intersection, a linear combination of the $U(1)_1\times U(1)_2$, which we call $U(1)_C$, will have mixed anomalies. The gauge symmetry $U(1)_C$ and the chiral matter living at the intersection\footnote{In type IIB literature, the phases $b$ are usually called \textit{open string axions} \cite{Cicoli:2012sz,Cicoli:2013ana}.} play the role of the ingredients described  previously in section \ref{section2}. 

The anomalies of $U(1)_C$ are canceled by an anomaly inflow mechanism associated to the RR 4-form $C_4$ which is, in some sense, a generalisation of the Green-Schwarz mechanism~\cite{Aldazabal:2000dg}, where the $C_4$ shifts under the anomalous $U(1)_C$. This can be seen from the action
\begin{equation}\label{eq:action_typeIIB}
S\supset \kappa_{1}  \int_{D7_{1}} C_4\wedge F_{1}\wedge F_{1}+\kappa_{2}  \int_{D7_{2}} C_4\wedge F_{2}\wedge F_{2}+S_{D7_1 \cap D7_2}\,.
\end{equation}
where a localised gauge anomaly at the brane intersection can be cancelled with the appropriate coefficients~$\kappa_{1,2}$. 

This can also be understood in terms of 4d axion fields and their transformation under the $U(1)_C$. There are two independent axions arising from integrating the RR 4-form field over $W_{1,2}$,
\begin{equation}
    \theta_{1,\,2}=\int_{W_{1,2}}C_4\,.
\end{equation}
By construction, to cancel the localised anomaly, $C_4$ has to shift appropriately under $U(1)_C$ so that the action in Eq.~\eqref{eq:action_typeIIB} is gauge invariant. In this case, there will be two un-eaten gauge invariant linear combinations  involving $\theta_1$, $\theta_2$, and the phase of the complex scalar, $b$ which are higher axions.

We now come to the issue of scale separation. In minimal models with pseudo-anomalous $U(1)$s, the VEV of the complex scalar lies close to the string scale, which also determines the mass of the anomalous gauge boson. Similar to the heterotic string case, this occurs through the $D-$term which has a contribution from $\Phi$ as well as the moduli-dependent FI terms. In the case of type IIB with $D7$-branes, the $D-$term relevant for $U(1)_C$ is
\begin{equation}
    D_C= -q_C|\Phi|^2+\xi \,.
\end{equation}
where $\xi$ is a function of the moduli fields. For the toy model with $D7-$branes wrapping $W_1$ and $W_2$, we have
\begin{equation}\label{eq:FI-typeIIB}
    \xi = q_1\frac{\partial K}{\partial \tau_1} + q_2\frac{\partial K}{\partial \tau_2}\,,
\end{equation}
where $q_C$, $q_{1}$, $q_2$ are the associated $U(1)_C$ charges. While we do not present any concrete CY, it is conceivable to find special loci in moduli space where $\xi$ vanishes, in close analogy to the heterotic case~\cite{Buchbinder:2014qca}. Near this region, the VEV $|\Phi|^2\propto \xi$ can be made small.  In explicit constructions one should also add the contribution from the dilaton and any other RR field which shifts under $U(1)_C$. In addition, in order to solve the strong CP problem and be detectable through their coupling to SM gauge bosons, these constructions should include chiral matter charged under the SM gauge sector at the intersection. We leave explicit realisations for future work.

We remark that the size of the decay constant of the higher axion, $F_\theta\simeq |\Phi|$, is small near de loci due to a non-trivial cancellation between contributions to the $D-$term instead of due to the existence of a collapsing cycle (see \cite{Cicoli:2013ana} for examples of this kind). Hence, we do not expect to have unsuppressed instantons spoiling the quality of $\theta$ because this kind of cancellations are not expected to appear at the level of the $D-$instanton action.

\section{Conclusion}
\label{discussion}
In this work we presented a new mechanism to obtain field theoretic axion strings for exponentially good quality axions. The minimal version of the mechanism -- presented in the form of a 5d theory with boundary -- involves the mixing between an axion coming from the fifth component of a bulk gauge field and the phase of a complex scalar that lives on the boundary. This differs qualitatively from standard field theory as well as higher-dimensional axions, and resembles situations that appear in the context of higher-group symmetries. For this reason, we named the 4d axion $\theta$ as \textit{higher axion}. Crucially, higher axion strings can be produced by standard Kibble-Zurek mechanism in the early universe and constitutes a natural way to reconcile the predictivity of the post-inflationary axion scenario together with the attractive features of the string axiverse. 

We have shown that the key ingredients are common in most of string constructions, providing guidance to obtain higher axions in heterotic strings as well as in type II compactifications. As usual in scenarios with an anomalous $U(1)$, FI terms play a relevant role. In field theory, one has in principle freedom to choose the VEV of the complex scalar with respect to the compactification scale. However, in supersymmetric string compactifications, these two scales are tied through the $D-$term potential. We have shown that it is in principle possible to separate $|\Phi|$ from the string scale, $M_s$, if multiple moduli fields shift under the same anomalous $U(1)$. This occurs through cancellations between different contributions to the FI term, which are exact in the certain regions of moduli space.   

For the case of the QCD axion, numerical simulations of the post-inflationary scenario seem to point towards a low axion decay constant around $F_\theta\sim 10^{10}-10^{11}$ GeV. In the case of heterotic string theory, where the string scale is typically around $M_s\sim 10^{17}$ GeV, this seems to require a fine cancellation between different FI term contributions (see Eq.~\eqref{eq:D-term_heterotic}). On the other hand, in type IIB scenarios, the situation may be improved thanks to the freedom in decreasing the string scale with respect to the 4d Planck scale by going to the large volume limit~\cite{Conlon:2006tq}. Intuitively, in these scenarios one lowers $F_\theta$ by lowering the string scale, since both quantities decrease as the total volume of the CY increases. Therefore, as shown in Eq.~\eqref{eq:FI-typeIIB}, one can get smaller $F_\theta$ with milder cancellations. The required cancellation being milder is equivalent to having larger regions in moduli space where the higher axion decay constant is close to the required values $F_\theta\sim 10^{10}-10^{11}$ GeV.

Finally, we note that moduli stabilisation may have a large impact in the cancellations between different contributions to the FI terms. Understanding their contributions and finding explicitly string theory constructions of the mechanisms presented here are very interesting directions that we leave for future work. 
\\\\
\textbf{\textit{Note added:}} while we were finishing the writing of this manuscript Ref.~\cite{Petrossian-Byrne:2025mto} appeared partially overlapping with the result of this work and reaching similar conclusions. 

\acknowledgments
We would like to thank Michele Cicoli, Joe Conlon, Thomas Harvey, Arthur Hebecker, John March-Russell, André Lukas and Lucas Leung  for useful discussions. MR thanks Naomi Gendler, Arthur Hebecker, John March-Russell, Doddy Marsh, Jakob Moritz, Liam McAllister and other participants of the \hyperlink{http://www.birs.ca/events/2025/5-day-workshops/25w5384}{Banff Axiverse workshop} for discussions on these and related topics. 
We are indebted to Prateek Agrawal for his feedback, advice, and many useful discussions on this and related projects -- without these interactions, this project would never have occurred.
AP is supported by a STFC Studenship No. 2397217 and Cultuurfondsbeurs No. 40038041 made possible by the Pieter Beijer fonds and the Data-Piet fonds.
VL is supported by the STFC under Grant No.~ST/X000761/1.

\bibliography{ref,newrefs_axion-6}

\providecommand{\href}[2]{#2}\begingroup\raggedright\begin{thebibliography}{10}

\bibitem{Graham:2015ouw}
P.~W. Graham, I.~G. Irastorza, S.~K. Lamoreaux, A.~Lindner, and K.~A. van
  Bibber, {\it {Experimental Searches for the Axion and Axion-Like Particles}},
   {\em Ann. Rev. Nucl. Part. Sci.} {\bf 65} (2015) 485--514,
  [\href{http://arxiv.org/abs/1602.00039}{{\tt arXiv:1602.00039}}].

\bibitem{Irastorza:2018dyq}
I.~G. Irastorza and J.~Redondo, {\it {New experimental approaches in the search
  for axion-like particles}},  {\em Prog. Part. Nucl. Phys.} {\bf 102} (2018)
  89--159, [\href{http://arxiv.org/abs/1801.08127}{{\tt arXiv:1801.08127}}].

\bibitem{Weinberg:1977ma}
S.~Weinberg, {\it {A New Light Boson?}},  {\em Phys. Rev. Lett.} {\bf 40}
  (1978) 223--226.

\bibitem{Wilczek:1977pj}
F.~Wilczek, {\it {Problem of Strong $P$ and $T$ Invariance in the Presence of
  Instantons}},  {\em Phys. Rev. Lett.} {\bf 40} (1978) 279--282.

\bibitem{Peccei:1977hh}
R.~D. Peccei and H.~R. Quinn, {\it {CP Conservation in the Presence of
  Instantons}},  {\em Phys. Rev. Lett.} {\bf 38} (1977) 1440--1443.

\bibitem{Preskill:1982cy}
J.~Preskill, M.~B. Wise, and F.~Wilczek, {\it {Cosmology of the Invisible
  Axion}},  {\em Phys. Lett. B} {\bf 120} (1983) 127--132.

\bibitem{Dine:1982ah}
M.~Dine and W.~Fischler, {\it {The Not So Harmless Axion}},  {\em Phys. Lett.
  B} {\bf 120} (1983) 137--141.

\bibitem{Abbott:1982af}
L.~F. Abbott and P.~Sikivie, {\it {A Cosmological Bound on the Invisible
  Axion}},  {\em Phys. Lett. B} {\bf 120} (1983) 133--136.

\bibitem{Svrcek:2006yi}
P.~Svrcek and E.~Witten, {\it {Axions In String Theory}},  {\em JHEP} {\bf 06}
  (2006) 051, [\href{http://arxiv.org/abs/hep-th/0605206}{{\tt
  hep-th/0605206}}].

\bibitem{Arvanitaki:2009fg}
A.~Arvanitaki, S.~Dimopoulos, S.~Dubovsky, N.~Kaloper, and J.~March-Russell,
  {\it {String Axiverse}},  {\em Phys. Rev. D} {\bf 81} (2010) 123530,
  [\href{http://arxiv.org/abs/0905.4720}{{\tt arXiv:0905.4720}}].

\bibitem{Borsanyi:2016ksw}
S.~Borsanyi et~al., {\it {Calculation of the axion mass based on
  high-temperature lattice quantum chromodynamics}},  {\em Nature} {\bf 539}
  (2016), no.~7627 69--71, [\href{http://arxiv.org/abs/1606.07494}{{\tt
  arXiv:1606.07494}}].

\bibitem{Vaquero:2018tib}
A.~Vaquero, J.~Redondo, and J.~Stadler, {\it {Early seeds of axion
  miniclusters}},  {\em JCAP} {\bf 04} (2019) 012,
  [\href{http://arxiv.org/abs/1809.09241}{{\tt arXiv:1809.09241}}].

\bibitem{Gorghetto:2018myk}
M.~Gorghetto, E.~Hardy, and G.~Villadoro, {\it {Axions from Strings: the
  Attractive Solution}},  {\em JHEP} {\bf 07} (2018) 151,
  [\href{http://arxiv.org/abs/1806.04677}{{\tt arXiv:1806.04677}}].

\bibitem{Buschmann:2019icd}
M.~Buschmann, J.~W. Foster, and B.~R. Safdi, {\it {Early-Universe Simulations
  of the Cosmological Axion}},  {\em Phys. Rev. Lett.} {\bf 124} (2020), no.~16
  161103, [\href{http://arxiv.org/abs/1906.00967}{{\tt arXiv:1906.00967}}].

\bibitem{Gorghetto:2020qws}
M.~Gorghetto, E.~Hardy, and G.~Villadoro, {\it {More Axions from Strings}},
  \href{http://arxiv.org/abs/2007.04990}{{\tt arXiv:2007.04990}}.

\bibitem{Kim:2024dtq}
H.~Kim and M.~Son, {\it {More Scalings from Cosmic Strings}},
  \href{http://arxiv.org/abs/2411.08455}{{\tt arXiv:2411.08455}}.

\bibitem{Saikawa:2024bta}
K.~Saikawa, J.~Redondo, A.~Vaquero, and M.~Kaltschmidt, {\it {Spectrum of
  global string networks and the axion dark matter mass}},  {\em JCAP} {\bf 10}
  (2024) 043, [\href{http://arxiv.org/abs/2401.17253}{{\tt arXiv:2401.17253}}].

\bibitem{Correia:2024cpk}
J.~Correia, M.~Hindmarsh, J.~Lizarraga, A.~Lopez-Eiguren, K.~Rummukainen, and
  J.~Urrestilla, {\it {Scaling density of axion strings in terasite
  simulations}},  {\em Phys. Rev. D} {\bf 111} (2025), no.~6 063532,
  [\href{http://arxiv.org/abs/2410.18064}{{\tt arXiv:2410.18064}}].

\bibitem{Benabou:2024msj}
J.~N. Benabou, M.~Buschmann, J.~W. Foster, and B.~R. Safdi, {\it {Axion mass
  prediction from adaptive mesh refinement cosmological lattice simulations}},
  \href{http://arxiv.org/abs/2412.08699}{{\tt arXiv:2412.08699}}.

\bibitem{Kamionkowski:1992mf}
M.~Kamionkowski and J.~March-Russell, {\it {Planck scale physics and the
  Peccei-Quinn mechanism}},  {\em Phys. Lett.} {\bf B282} (1992) 137--141,
  [\href{http://arxiv.org/abs/hep-th/9202003}{{\tt hep-th/9202003}}].

\bibitem{Kallosh:1995hi}
R.~Kallosh, A.~D. Linde, D.~A. Linde, and L.~Susskind, {\it {Gravity and global
  symmetries}},  {\em Phys. Rev. D} {\bf 52} (1995) 912--935,
  [\href{http://arxiv.org/abs/hep-th/9502069}{{\tt hep-th/9502069}}].

\bibitem{Dine:2022mjw}
M.~Dine, {\it {The Problem of Axion Quality: A Low Energy Effective Action
  Perspective}},  \href{http://arxiv.org/abs/2207.01068}{{\tt
  arXiv:2207.01068}}.

\bibitem{Chun:1992bn}
E.~J. Chun and A.~Lukas, {\it {Discrete gauge symmetries in axionic extensions
  of the SSM}},  {\em Phys. Lett. B} {\bf 297} (1992) 298--304,
  [\href{http://arxiv.org/abs/hep-ph/9209208}{{\tt hep-ph/9209208}}].

\bibitem{Ringwald:2015dsf}
A.~Ringwald and K.~Saikawa, {\it {Axion dark matter in the post-inflationary
  Peccei-Quinn symmetry breaking scenario}},  {\em Phys. Rev. D} {\bf 93}
  (2016), no.~8 085031, [\href{http://arxiv.org/abs/1512.06436}{{\tt
  arXiv:1512.06436}}]. [Addendum: Phys.Rev.D 94, 049908 (2016)].

\bibitem{Fukuda:2017ylt}
H.~Fukuda, M.~Ibe, M.~Suzuki, and T.~T. Yanagida, {\it {A ''gauged'' $U(1)$
  Peccei\textendash{}Quinn symmetry}},  {\em Phys. Lett. B} {\bf 771} (2017)
  327--331, [\href{http://arxiv.org/abs/1703.01112}{{\tt arXiv:1703.01112}}].

\bibitem{Randall:1992ut}
L.~Randall, {\it {Composite axion models and Planck scale physics}},  {\em
  Phys. Lett. B} {\bf 284} (1992) 77--80.

\bibitem{Lillard:2017cwx}
B.~Lillard and T.~M.~P. Tait, {\it {A Composite Axion from a Supersymmetric
  Product Group}},  {\em JHEP} {\bf 11} (2017) 005,
  [\href{http://arxiv.org/abs/1707.04261}{{\tt arXiv:1707.04261}}].

\bibitem{Lillard:2018fdt}
B.~Lillard and T.~M.~P. Tait, {\it {A High Quality Composite Axion}},  {\em
  JHEP} {\bf 11} (2018) 199, [\href{http://arxiv.org/abs/1811.03089}{{\tt
  arXiv:1811.03089}}].

\bibitem{DiLuzio:2017tjx}
L.~Di~Luzio, E.~Nardi, and L.~Ubaldi, {\it {Accidental Peccei-Quinn symmetry
  protected to arbitrary order}},  {\em Phys. Rev. Lett.} {\bf 119} (2017),
  no.~1 011801, [\href{http://arxiv.org/abs/1704.01122}{{\tt
  arXiv:1704.01122}}].

\bibitem{Gavela:2018paw}
M.~B. Gavela, M.~Ibe, P.~Quilez, and T.~T. Yanagida, {\it {Automatic
  Peccei\textendash{}Quinn symmetry}},  {\em Eur. Phys. J. C} {\bf 79} (2019),
  no.~6 542, [\href{http://arxiv.org/abs/1812.08174}{{\tt arXiv:1812.08174}}].

\bibitem{Contino:2021ayn}
R.~Contino, A.~Podo, and F.~Revello, {\it {Chiral models of composite axions
  and accidental Peccei-Quinn symmetry}},  {\em JHEP} {\bf 04} (2022) 180,
  [\href{http://arxiv.org/abs/2112.09635}{{\tt arXiv:2112.09635}}].

\bibitem{Vecchi:2021shj}
L.~Vecchi, {\it {Axion quality straight from the GUT}},  {\em Eur. Phys. J. C}
  {\bf 81} (2021), no.~10 938, [\href{http://arxiv.org/abs/2106.15224}{{\tt
  arXiv:2106.15224}}].

\bibitem{Choi:2022fha}
G.~Choi and T.~T. Yanagida, {\it {High quality axion in supersymmetric
  models}},  {\em JHEP} {\bf 12} (2022) 067,
  [\href{http://arxiv.org/abs/2209.09290}{{\tt arXiv:2209.09290}}].

\bibitem{Babu:2024udi}
K.~S. Babu, B.~Dutta, and R.~N. Mohapatra, {\it {Hybrid SO(10) Axion Model
  Without Quality Problem}},  \href{http://arxiv.org/abs/2410.07323}{{\tt
  arXiv:2410.07323}}.

\bibitem{Babu:2024qzb}
K.~S. Babu, B.~Dutta, and R.~N. Mohapatra, {\it {Accidental Peccei-Quinn
  Symmetry From Gauged U(1) and a High Quality Axion}},
  \href{http://arxiv.org/abs/2412.21157}{{\tt arXiv:2412.21157}}.

\bibitem{Arkani-Hamed:1998jmv}
N.~Arkani-Hamed, S.~Dimopoulos, and G.~R. Dvali, {\it {The Hierarchy problem
  and new dimensions at a millimeter}},  {\em Phys. Lett. B} {\bf 429} (1998)
  263--272, [\href{http://arxiv.org/abs/hep-ph/9803315}{{\tt hep-ph/9803315}}].

\bibitem{Randall:1999ee}
L.~Randall and R.~Sundrum, {\it {A Large mass hierarchy from a small extra
  dimension}},  {\em Phys. Rev. Lett.} {\bf 83} (1999) 3370--3373,
  [\href{http://arxiv.org/abs/hep-ph/9905221}{{\tt hep-ph/9905221}}].

\bibitem{Craig:2024dnl}
N.~Craig and M.~Kongsore, {\it {High-quality axions from higher-form symmetries
  in extra dimensions}},  {\em Phys. Rev. D} {\bf 111} (2025), no.~1 015047,
  [\href{http://arxiv.org/abs/2408.10295}{{\tt arXiv:2408.10295}}].

\bibitem{Agrawal:2024ejr}
P.~Agrawal, M.~Nee, and M.~Reig, {\it {Axion couplings in heterotic string
  theory}},  {\em JHEP} {\bf 02} (2025) 188,
  [\href{http://arxiv.org/abs/2410.03820}{{\tt arXiv:2410.03820}}].

\bibitem{Reece:2024wrn}
M.~Reece, {\it {Extra-Dimensional Axion Expectations}},
  \href{http://arxiv.org/abs/2406.08543}{{\tt arXiv:2406.08543}}.

\bibitem{Benabou:2023npn}
J.~N. Benabou, Q.~Bonnefoy, M.~Buschmann, S.~Kumar, and B.~R. Safdi, {\it
  {Cosmological dynamics of string theory axion strings}},  {\em Phys. Rev. D}
  {\bf 110} (2024), no.~3 035021, [\href{http://arxiv.org/abs/2312.08425}{{\tt
  arXiv:2312.08425}}].

\bibitem{Cheng:2001ys}
H.-C. Cheng and D.~E. Kaplan, {\it {Axions and a gauged Peccei-Quinn
  symmetry}},  \href{http://arxiv.org/abs/hep-ph/0103346}{{\tt
  hep-ph/0103346}}.

\bibitem{Cordova:2018cvg}
C.~C\'ordova, T.~T. Dumitrescu, and K.~Intriligator, {\it {Exploring 2-Group
  Global Symmetries}},  {\em JHEP} {\bf 02} (2019) 184,
  [\href{http://arxiv.org/abs/1802.04790}{{\tt arXiv:1802.04790}}].

\bibitem{March-Russell:2021zfq}
J.~March-Russell and H.~Tillim, {\it {Axiverse Strings}},
  \href{http://arxiv.org/abs/2109.14637}{{\tt arXiv:2109.14637}}.

\bibitem{Cline:2024vbd}
J.~M. Cline, C.~Litos, and W.~Xue, {\it {Axion strings from string axions}},
  \href{http://arxiv.org/abs/2412.12260}{{\tt arXiv:2412.12260}}.

\bibitem{Kibble:1976sj}
T.~W.~B. Kibble, {\it {Topology of Cosmic Domains and Strings}},  {\em J. Phys.
  A} {\bf 9} (1976) 1387--1398.

\bibitem{Kawamura:1999nj}
Y.~Kawamura, {\it {Gauge symmetry breaking from extra space S**1 / Z(2)}},
  {\em Prog. Theor. Phys.} {\bf 103} (2000) 613--619,
  [\href{http://arxiv.org/abs/hep-ph/9902423}{{\tt hep-ph/9902423}}].

\bibitem{Bhardwaj:2023kri}
L.~Bhardwaj, L.~E. Bottini, L.~Fraser-Taliente, L.~Gladden, D.~S.~W. Gould,
  A.~Platschorre, and H.~Tillim, {\it {Lectures on generalized symmetries}},
  {\em Phys. Rept.} {\bf 1051} (2024) 1--87,
  [\href{http://arxiv.org/abs/2307.07547}{{\tt arXiv:2307.07547}}].

\bibitem{Green:1984sg}
M.~B. Green and J.~H. Schwarz, {\it {Anomaly Cancellation in Supersymmetric
  D=10 Gauge Theory and Superstring Theory}},  {\em Phys. Lett. B} {\bf 149}
  (1984) 117--122.

\bibitem{Fraser:2019ojt}
K.~Fraser and M.~Reece, {\it {Axion Periodicity and Coupling Quantization in
  the Presence of Mixing}},  {\em JHEP} {\bf 05} (2020) 066,
  [\href{http://arxiv.org/abs/1910.11349}{{\tt arXiv:1910.11349}}].

\bibitem{Caputo:2024oqc}
A.~Caputo and G.~Raffelt, {\it {Astrophysical Axion Bounds: The 2024 Edition}},
   {\em PoS} {\bf COSMICWISPers} (2024) 041,
  [\href{http://arxiv.org/abs/2401.13728}{{\tt arXiv:2401.13728}}].

\bibitem{Hook:2018dlk}
A.~Hook, {\it {TASI Lectures on the Strong CP Problem and Axions}},  {\em PoS}
  {\bf TASI2018} (2019) 004, [\href{http://arxiv.org/abs/1812.02669}{{\tt
  arXiv:1812.02669}}].

\bibitem{Choi:2003wr}
K.-w. Choi, {\it {A QCD axion from higher dimensional gauge field}},  {\em
  Phys. Rev. Lett.} {\bf 92} (2004) 101602,
  [\href{http://arxiv.org/abs/hep-ph/0308024}{{\tt hep-ph/0308024}}].

\bibitem{Horava:1996ma}
P.~Horava and E.~Witten, {\it {Eleven-dimensional supergravity on a manifold
  with boundary}},  {\em Nucl. Phys. B} {\bf 475} (1996) 94--114,
  [\href{http://arxiv.org/abs/hep-th/9603142}{{\tt hep-th/9603142}}].

\bibitem{Niu:2023khv}
X.~Niu, W.~Xue, and F.~Yang, {\it {Gauged global strings}},  {\em JHEP} {\bf
  02} (2024) 093, [\href{http://arxiv.org/abs/2311.07639}{{\tt
  arXiv:2311.07639}}].

\bibitem{Polchinski:1998rr}
J.~Polchinski, {\em {String theory. Vol. 2: Superstring theory and beyond}}.
\newblock Cambridge Monographs on Mathematical Physics. Cambridge University
  Press, 12, 2007.

\bibitem{Ibanez:2012zz}
L.~E. Ibanez and A.~M. Uranga, {\em {String theory and particle physics: An
  introduction to string phenomenology}}.
\newblock Cambridge University Press, 2, 2012.

\bibitem{Candelas:1985en}
P.~Candelas, G.~T. Horowitz, A.~Strominger, and E.~Witten, {\it {Vacuum
  configurations for superstrings}},  {\em Nucl. Phys. B} {\bf 258} (1985)
  46--74.

\bibitem{Dine:1987xk}
M.~Dine, N.~Seiberg, and E.~Witten, {\it {Fayet-Iliopoulos Terms in String
  Theory}},  {\em Nucl. Phys. B} {\bf 289} (1987) 589--598.

\bibitem{Gross:1984dd}
D.~J. Gross, J.~A. Harvey, E.~J. Martinec, and R.~Rohm, {\it {The Heterotic
  String}},  {\em Phys. Rev. Lett.} {\bf 54} (1985) 502--505.

\bibitem{Blumenhagen:2005ga}
R.~Blumenhagen, G.~Honecker, and T.~Weigand, {\it {Loop-corrected
  compactifications of the heterotic string with line bundles}},  {\em JHEP}
  {\bf 06} (2005) 020, [\href{http://arxiv.org/abs/hep-th/0504232}{{\tt
  hep-th/0504232}}].

\bibitem{Anderson:2011ns}
L.~B. Anderson, J.~Gray, A.~Lukas, and E.~Palti, {\it {Two Hundred Heterotic
  Standard Models on Smooth Calabi-Yau Threefolds}},  {\em Phys. Rev. D} {\bf
  84} (2011) 106005, [\href{http://arxiv.org/abs/1106.4804}{{\tt
  arXiv:1106.4804}}].

\bibitem{Anderson:2012yf}
L.~B. Anderson, J.~Gray, A.~Lukas, and E.~Palti, {\it {Heterotic Line Bundle
  Standard Models}},  {\em JHEP} {\bf 06} (2012) 113,
  [\href{http://arxiv.org/abs/1202.1757}{{\tt arXiv:1202.1757}}].

\bibitem{Buchbinder:2014qca}
E.~I. Buchbinder, A.~Constantin, and A.~Lukas, {\it {Heterotic QCD axion}},
  {\em Phys. Rev. D} {\bf 91} (2015), no.~4 046010,
  [\href{http://arxiv.org/abs/1412.8696}{{\tt arXiv:1412.8696}}].

\bibitem{Im:2019cnl}
S.~H. Im, H.~P. Nilles, and M.~Olechowski, {\it {Axion clockworks from
  heterotic M-theory: the QCD-axion and its ultra-light companion}},  {\em
  JHEP} {\bf 10} (2019) 159, [\href{http://arxiv.org/abs/1906.11851}{{\tt
  arXiv:1906.11851}}].

\bibitem{Blesneag:2015pvz}
S.~Blesneag, E.~I. Buchbinder, P.~Candelas, and A.~Lukas, {\it {Holomorphic
  Yukawa Couplings in Heterotic String Theory}},  {\em JHEP} {\bf 01} (2016)
  152, [\href{http://arxiv.org/abs/1512.05322}{{\tt arXiv:1512.05322}}].

\bibitem{Douglas:2006hz}
M.~R. Douglas, R.~L. Karp, S.~Lukic, and R.~Reinbacher, {\it {Numerical
  solution to the hermitian Yang-Mills equation on the Fermat quintic}},  {\em
  JHEP} {\bf 12} (2007) 083, [\href{http://arxiv.org/abs/hep-th/0606261}{{\tt
  hep-th/0606261}}].

\bibitem{Harvey:2021oue}
T.~R. Harvey and A.~Lukas, {\it {Quark Mass Models and Reinforcement
  Learning}},  {\em JHEP} {\bf 08} (2021) 161,
  [\href{http://arxiv.org/abs/2103.04759}{{\tt arXiv:2103.04759}}].

\bibitem{Constantin:2024yxh}
A.~Constantin, C.~S. Fraser-Taliente, T.~R. Harvey, A.~Lukas, and B.~Ovrut,
  {\it {Computation of quark masses from string theory}},  {\em Nucl. Phys. B}
  {\bf 1010} (2025) 116778, [\href{http://arxiv.org/abs/2402.01615}{{\tt
  arXiv:2402.01615}}].

\bibitem{Uranga:933469}
A.~M. Uranga, {\it {TASI lectures on string compactification, model building
  and fluxes}},  tech. rep., CERN, Geneva, 2005.

\bibitem{Cicoli:2012sz}
M.~Cicoli, M.~Goodsell, and A.~Ringwald, {\it {The type IIB string axiverse and
  its low-energy phenomenology}},  {\em JHEP} {\bf 10} (2012) 146,
  [\href{http://arxiv.org/abs/1206.0819}{{\tt arXiv:1206.0819}}].

\bibitem{Cicoli:2013ana}
M.~Cicoli, {\it {Axion-like Particles from String Compactifications}},  in {\em
  {9th Patras Workshop on Axions, WIMPs and WISPs}}, pp.~235--242, 2013.
\newblock \href{http://arxiv.org/abs/1309.6988}{{\tt arXiv:1309.6988}}.

\bibitem{Aldazabal:2000dg}
G.~Aldazabal, S.~Franco, L.~E. Ibanez, R.~Rabadan, and A.~M. Uranga, {\it {D =
  4 chiral string compactifications from intersecting branes}},  {\em J. Math.
  Phys.} {\bf 42} (2001) 3103--3126,
  [\href{http://arxiv.org/abs/hep-th/0011073}{{\tt hep-th/0011073}}].

\bibitem{Conlon:2006tq}
J.~P. Conlon, {\it {The QCD axion and moduli stabilisation}},  {\em JHEP} {\bf
  05} (2006) 078, [\href{http://arxiv.org/abs/hep-th/0602233}{{\tt
  hep-th/0602233}}].

\bibitem{Petrossian-Byrne:2025mto}
R.~Petrossian-Byrne and G.~Villadoro, {\it {Open String Axiverse}},
  \href{http://arxiv.org/abs/2503.16387}{{\tt arXiv:2503.16387}}.

\end{thebibliography}\endgroup
%\nocite{*}
\bibliographystyle{JHEP}

\end{document}